# Effective Scalar Theory of the Electroweak Phase Transition

F. Karsch[a], T. Neuhaus[b] and A. Patkós[c]

[a]Fakultät für Physik, Universität Bielefeld,
D-33615 Bielefeld, F.R. Germany,

[b]FB8 Physik, BUGH Wuppertal, F.R. Germany

[c]Dept. of Atomic Physics, Eötvös University,
H-1088 Puskin u. 5-7., Hungary

This talk summarizes results obtained from simulating a three dimensional pure Order Parameter (OP) model of the Electroweak Phase Transition (EWPT). Its detailed presentation appears in [1].

## 1. Introduction

The key to the electroweak phase transition lies in proper understanding of the interaction of static ($n=0$ Matsubara) modes of the Standard Model. In particular, the correct interpretation of the way magnetic vector fluctuations drive the breakdwon of the $O(4)$ symmetry into a first order transition is of central importance.

The optimistic approach tends to limit the role of vector fluctuations to induce a "cubic" term into the potential energy of the order parameter (Higgs) fields. This situation arises already upon 1-loop integration over the static $A_i^a(x)$ fields.

This view might be substantiated by investigations of the $SU(2)$ Higgs model with infinitely many Higgs doublets. With appropriate large N scaling of the couplings of the model one can show that the gauge fields $A_0(x), A_i(x)$ can be integrated out exactly, and a pure OP-theory is arrived at containing the peculiar "cubic" piece in its potential energy.

Another way of looking for the range of couplings, where pure OP-theory might be the adequate description, is to inspect the thermal mass hierarchy of different fluctuations near $T_c$, as calculated with improved perturbative tools. In the region of small ($\leq 50$ GeV) Higgs masses, the thermal mass of the scalars is much below the Debye-mass of $A_0$, and is reasonably smaller of $A_i$, too [2].

The main argument against this extreme reducibility is that non-perturbative effects, like condensate of 3-dimensional, static vector quanta might be missed [3].

This criticism of the perturbative approach might have contributed to the expansion of numerical simulations of EWPT, experienced during the last year. Large scale studies of the full finite temperature system receive renewed interest [4,5]. The range of the Higgs mass explored to date is below 50 GeV, where the first order signal is strong.

In simulations of the reduced (static) sector of the model [6,7] careful implementation of the curves of "equal physics" is the most sensitive part of the interpretation of lattice data.

## 2. Continuum limit of the lattice model

Our goal with the present investigation was to study effects exerted by scalar fluctuations on the transition. Magnetic and electric field fluctuations were treated perturbatively on the 1-loop level. The discrete lattice action of the model is as follows:

$$S_{3d,lat} = \sum_x \left[ \frac{1}{2\kappa} \psi_x^+ \psi_x + \frac{\hat{\lambda}\Theta}{24}(\psi_x^+ \psi_x)^2 \right.$$
$$-\frac{1}{2}\sum_e (\psi_{x+e}^+ \psi_x + \psi_x^+ \psi_{x+e}) - \frac{g^3\Theta^{3/2}}{32\pi} \times$$
$$\left. (2(\gamma_M\Theta + \psi_x^+\psi_x)^{3/2} + (\gamma_E\Theta + \psi_x^+\psi_x)^{3/2}) \right] \quad (1)$$

with $\psi_x$ representing the Higgs-field in form of a 4-component site variable; $\Theta = aT$ is the temperature scaled by the lattice spacing: $a$. The gauge coupling was fixed by requiring the vacuum expectation value to be 246 GeV, and the W-mass $m_W=80$ GeV. The electric and magnetic mass parameters are

$$\gamma_E = \frac{10}{3}, \qquad \gamma_M = \frac{4g^2}{9\pi^4}. \qquad (2)$$

For the first quantity its perturbative value seems to be a rather good approximation, while for the value of the magnetic screening mass we used the selfconsistent solution of a corresponding Schwinger-Dyson equation [8,9].

The remaining 3 parameters ($\kappa$, $\hat\lambda$ and $\Theta$) of the lattice action (1) can be tuned simultanously to reach the continuum limit and $T_c$. The first two are related to the renormalised parameters of the continuum model $m^2$ and $\lambda$ through the 1-loop relations:

$$\frac{1}{2\kappa} = \frac{1}{2}\hat m^2 a^2 + \frac{1}{2}(\frac{3}{16}g^2 + \frac{\lambda}{12})\Theta^2 - \frac{C}{2}\Theta\Sigma(L^3) + 3 \ , (3)$$

$$\hat m^2 = m^2\{1 - \tfrac{1}{32\pi^2}[(\tfrac{9}{2}g^2 + \lambda)\ln\tfrac{3g^2 v_0^2}{4T^2} +\lambda\ln\tfrac{\lambda v_0^2}{3T^2}] - \tfrac{1}{128\pi^2}(45g^2 + 20\lambda + \tfrac{27g^4}{\lambda})\}, \quad (4)$$

$$\hat\lambda = \lambda - \tfrac{9}{16\pi^2}(\tfrac{9g^4}{16} + \tfrac{3g^2\lambda}{4} + \tfrac{\lambda^2}{3}) - \tfrac{3}{8\pi^2}\{g^4(\tfrac{3}{8}\ln\tfrac{g^2 v_0^2}{4T^2} - \tfrac{3}{2}\ln\tfrac{g^2 v_0^2}{\sqrt{2}T^2}) + \tfrac{\lambda^2}{4}\ln\tfrac{\lambda v_0^2}{3T^2} + 3(\tfrac{3g^2}{4} + \tfrac{\lambda}{6})^2\ln\tfrac{3g^2 v_0^2}{4T^2}\}. \qquad (5)$$

The complicated relations $\hat m^2(m^2, T), \hat\lambda(\lambda, T)$ arise from imposing Linde's renormalisation conditions at $T = 0$ [10]. When compared with other (simpler) renormalisation schemes significant differences were found for $m_H(T = 0) \leq 30 GeV$.

In order to ensure comparability with results from simulations of the more complete version of the 3-dimensional reduced model [6] we have chosen $\hat\lambda$ to give finally for $\lambda$ a value corresponding to $m_H \approx 35 GeV$.

$\kappa$ has been tuned at fixed $\Theta$ to the transition point indicated by the fast variation of the order parameters

$$O_1 = \tfrac{1}{L^3}\sum_x \psi_x^+\psi_x \ , \quad O_2 = \tfrac{1}{3L^3}\sum_{x,e} \psi_x^+\psi_{x+e} \ ,$$
$$O_3 = \tfrac{1}{L^3}\sqrt{\{(\sum_x \psi_x)^+(\sum_y \psi_y)\}}. \qquad (6)$$

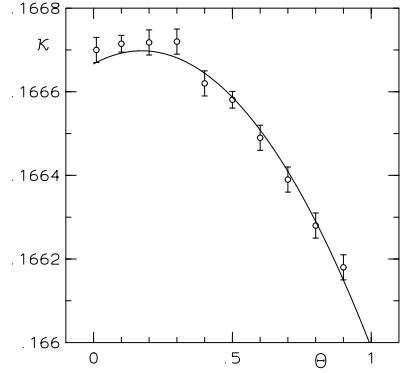

Figure 1. The critical line in the $\kappa - \Theta$ plane.

Finally, letting $\Theta$ to go to zero, one can reach the continuum limit.

The transition line in the $(\Theta, \kappa)$-plane is shown in Fig. 1, as it has been determined on a lattice of linear size $L = 18$. The solid line presents a least square fit to the data points. The solution of the equations determining the values of $T_c$ and $m_H$ is simpler if one replaces in the $\Theta \to 0$ limit (3) by

$$\lim_{\Theta \to 0} \tfrac{1}{\Theta^2}(\tfrac{1}{2\kappa} + \tfrac{C}{2}\Theta\Sigma(L^3) - 3)$$
$$\equiv Z_0 = \tfrac{\hat m^2}{2T_c^2} + \tfrac{1}{2}(\tfrac{3}{16}g^2 + \tfrac{\lambda}{12}). \qquad (7)$$

where $Z_{0,MC} = 0.0187(16)$ is a parameter which can be fitted from the solid curve of Fig. 1. Simultanous solution of (5) and (7) yields

$$T_c = 114.3(30) \ GeV, \quad m_H = 37.2 \ GeV. \qquad (8)$$

The value of $T_c$ was found to vary within 1 GeV, when different renormalisation conditions were imposed. Its value is higher than the corresponding prediction from the improved perturbative treatment (99.6 GeV). The value of $T_c$ found by [6] lies below the perturbative value!

## 3. Physical characteristics of the transition

Several physical quantities have been evaluated at fairly high values of $\Theta$ (=3,4,5) in order to enhance the signal. Below, we shortly review the most important results:

1. Accurate determination of $\kappa_c(\Theta)$, $\Theta = 3, 4, 5$ has been realised from finite size scaling (FSS) extrapolation, with inverse power correction in the size $L$, characteristic for first order transitions. Pseudocritical values of $\kappa$ were associated with maxima of $L^3 < O_i^2 - < O_i >^2 >$. $\mathcal{O}(10^{-4})$ accurate agreement was found with the Mean Field (MF) estimates.

2. Finite size estimates of the order parameter discontinuity have been extracted by measuring the distance of the peaks in the $< O_i >$ histograms, corresponding respectively to the symmetric (S) and broken symmetry (SB) phases. All 3 OP's gave mutually close infinite volume estimates. The physical value in proportion to $T_c$ is

$$\frac{\Delta \Phi}{T}|_{T=T_c} = 0.68(4), \quad (9)$$

a factor 2 smaller than the perturbative value and a factor 3 smaller than the numerical estimate of [6].

3. The latent heat has been estimated by substituting the numerically obtained value of the OP into the functional form of the internal energy. Forming the difference of the values taken in the symmetric and the broken symmetry phases one obtains the latent heat:

$$\frac{\mathcal{L}}{T_c^4} = 0.122(8), \quad (10)$$

what is again a factor of 2 smaller than the MF-estimate.

4. The interface tension between coexisting phases at $T_c$, has been measured by Binder's method. If in the volume of the lattice one has two well-separated domain walls, one can make use of the formula:

$$\sigma_{lat} = \frac{1}{2L^2} \ln \frac{P^{max}(O_i)}{P^{min}(O_i)} = \sigma_{phys} \frac{\Theta^2}{T^3}. \quad (11)$$

The separation of the walls has *not* been monitored in our runs.

For fixed $\Theta$ the slope of $\ln(P^{max}/P^{min})$ showed increasing tendency for increased lattice sizes. The average, therefore, might be rather a lower bound for $\sigma_{lat}(\Theta, L = \infty)$. With decreasing $\Theta$ an important drop (serious scaling violation) has been noticed in the value of the surface tension.

These contradicting tendencies do not allow to state more than an estimate of the surface tension, which proves to be surprisingly small:

$$\sigma_{MC} \sim \frac{1}{37} \sigma_{MF}. \quad (12)$$

4. Conclusion

Careful FSS investigations have established first order phase transition in the effective scalar field theory which has been obtained from the full finite temperature theory by hierarchical 1-loop integrations of all non-static and the static gauge degrees of freedom. All measured physical characteristics of the phase transition point uniformly to a weaker transition than predicted by the MF (perturbative) analysis. Since recent more complete simulations indicate within factor of 2 acceptable agreement between the results of the numerical simulation and of improved perturbation theory in the range 18 GeV $\leq m_H \leq$ 50 GeV, our results represent indirect hints for the importance of gauge field fluctuations in keeping the exact result close to the perturbative predictions.

REFERENCES

1. F. Karsch, T. Neuhaus and A. Patkós, Effective Scalar Field Theory for the Electroweak Phase Transition, BI-TP 94/27
2. A. Jakovác, K. Kajantie and A. Patkós, Phys. Rev. **D49** (1994) 6810
3. M. Shaposhnikov, Phys. Lett. **B316** (1993) 112
4. F. Csikor *et al.*, Phys. Lett. **B334** (1994) 405
5. Z. Fodor *et al.*, DESY-94-159 (Sept. 1994)
6. K. Kajantie, K. Rummukainen and M. Shaposhnikov, Nucl. Phys. **B407** (1993) 356
7. K. Farakos, K. Kajantie, K. Rummukainen and M. Shaposhnikov, Nucl. Phys. **B425** (1994) 67
8. W. Buchmüller, Z. Fodor, T. Helbig and D. Walliser, Ann. Phys. (N.Y.) **234** (1994) 260
9. J.R. Espinosa, M. Quirós and F. Zwirner, Phys. Lett. **B314** (1993) 206
10. A. Linde, Rep. Prog. Phys. **42** (1979) 389